\documentclass[a4paper,fleqn]{cas-sc}
\usepackage{amssymb}
\usepackage{amsmath}
\usepackage{xcolor}
\usepackage{enumitem}
\usepackage{xr}
\usepackage{hyperref}
\usepackage{multirow}
\usepackage{float}
\usepackage{caption}

\newcommand{\X}{\mathcal{X}}
\newcommand{\Real}{\mathbb{R}}
\newcommand{\Hl}{L_2(G)}
\newcommand{\mB}{\mathcal B}
\newcommand{\mZ}{\mathcal Z}
\newcommand{\Sd}{S^{\dag}}
\newcommand{\bt}{\beta}
\newcommand{\bth}{\widehat{\beta}}
\newcommand{\tht}{\theta}
\newcommand{\thth}{\widehat\theta}
\newcommand{\del}{\delta}
\newcommand{\delh}{\widehat\delta}
\newcommand{\sumk}{\sum_{i=1}^k}
\newcommand{\bts}{\beta^*}
\newcommand{\R}{\mathbb{R}}
\newcommand{\sig}{\sigma}
\newcommand{\sigh}{\widehat\sig}
\newcommand{\mE}{\mathbb{E}}
\newcommand{\N}{\mathcal{N}}
\newcommand{\al}{\alpha}
\newcommand{\als}{\alpha^*}
\newcommand{\alh}{\widehat{\alpha}}
\newcommand{\pconv}{\stackrel{\mathbb{P}}\longrightarrow}
\newcommand{\dconv}{\stackrel{d}\longrightarrow}
\newcommand{\pb}{\partial_{\beta}}
\newcommand{\nuh}{\widehat\nu}
\newcommand{\nut}{\widetilde{\nu}}
\newcommand{\pone}{\partial_1}
\newcommand{\ptwo}{\partial_2}
\newcommand{\etah}{\widehat{\eta}}
\newcommand{\mJ}{\mathcal J}
\newcommand{\mV}{\mathcal V}
\newcommand{\mC}{\mathcal C}
\newcommand{\btt}{\widetilde{\bt}}
\newcommand{\intmx}{\int_{\X}}

\newlist{assumption}{enumerate}{1}
\setlist[assumption]{label=\textbf{(A\arabic*)}, start=1, leftmargin=2em}

\usepackage[authoryear,longnamesfirst]{natbib}

\def\tsc#1{\csdef{#1}{\textsc{\lowercase{#1}}\xspace}}
\tsc{AB}
\tsc{SA}
\newtheorem{proposition}{Proposition}
\newtheorem{lemma}{Lemma}

\begin{document}
\let\WriteBookmarks\relax
\def\floatpagepagefraction{1}
\def\textpagefraction{.001}

% Short title
\shorttitle{}    

% Short author
\shortauthors{}  

\ExplSyntaxOn
\cs_gset:Npn \__first_footerline:
  {
    \group_begin:
    \small \sffamily
    \group_end:
  }
\ExplSyntaxOff

% Main title of the paper
\title [mode = title]{Compensator-based inference for signal detection under unknown background: the binned data case}  

\author[1]{Aritra Banerjee}
\cormark[1]
\ead{baner175@umn.edu}

\author[1]{Sara Algeri}[orcid=0000-0001-7366-3866]
\ead{salgeri@umn.edu}

\affiliation[1]{organization={School of Statistics, University of Minnesota},
            addressline={224 Church St SE}, 
            city={Minneapolis},
            postcode={55455}, 
            state={Minnesota},
            country={United States}}
\begin{abstract}
The problem of signal detection under an unknown background can be framed as one of inferring the weight of a mixture model with one misspecified component. \citet{banerjee_algeri_unbinned} show that, for this problem, the conservativeness of the inference is entirely determined by one single parameter, called \emph{the compensator}. They demonstrate that, when the data are independent and identically distributed, an inferential approach based on the compensator circumvents the need to estimate the density of the misspecified component and the associated challenges. The main purpose of this manuscript is to broaden the scope of such an approach and extend it to the case in which, as is often encountered in modern experiments in physics and astronomy, the data consist of Poisson counts observed over a large number of bins. 
\end{abstract}


\maketitle

\section{Introduction}\label{sec1}
Let $\X$ be a compact subset of $\Real^d$ and consider a  $d$-dimensional  grid on $\X$ composed of $k$ disjoint hyperrectangles $\{\Delta_i\}_{i=1}^k$ of equal volume $v = \frac{\|\X\|_d}{k}$, with $\|\cdot\|_d$ denoting the $d$-dimensional Euclidean norm, and centres $\{x_i\}_{i=1}^k$. Consider a collection $\{n_i\}_{i=1}^k$ of independent Poisson counts observed over these bins and such that:
\begin{equation}
\label{eqn:true_model}
E[n_i]=Tf_i\quad\text{with}\quad f_i=\int_{\Delta_i}f(x,\eta)dx\quad\text{and}\quad f(x,\eta)=(1-\eta)f_b(x)+\eta f_s(x)
\end{equation}
in which $f$ is assumed to be a probability density function in $x\in \mathcal{X}$ for all $\eta\in[0,1)$, and, letting $N=\sum_{i=1}^kn_i$, we set $T=E[N]$. The mixing parameter $\eta$ is the (unknown) relative signal intensity we aim to infer. Since some background is always present in all physics searches, it is assumed to be bounded away from one. 

The goal is to estimate $\eta$ and  test: 
\begin{equation}
\label{eqn:eta_test}H_0:\eta=0\quad\text{versus} \quad H_1:\eta>0\end{equation}
when $f_b$ is unknown and $f_s$ is known. Assuming $f_s$ to be known is a reasonable assumption since the nature of the signal is often well described by existing theories. Nevertheless, extensions to the situations in which such a density depends on free parameters are possible \citep[cf.][]{banerjee_algeri_unbinned}.

To best represent the nature of the data collected by modern experiments in physics and astronomy \citep[cf.][Sec. 1]{algeri_khmaladze}, the asymptotic regime considered is described by the following assumption:
\begin{equation}
\label{eqn:sparsness}
\frac{T}{k}\rightarrow c\in(0,\infty),\quad \text{as} \quad T,k\rightarrow \infty;
\end{equation}
that is, different from the case in which $k$ is fixed and $T$ grows without limit, the counts $\{n_i\}_{i=1}^k$ stay Poisson, and do not reach the Gaussian limit. 

Most solutions proposed in the literature to test \eqref{eqn:eta_test} \citep[e.g.,][]{atlas_spurious_signal,safeguard,bkg_modeling,algeri2020detecting,algeri2021informative, zhang_alageri_2023} estimate $f_b$ on a labeled dataset, known as the `background-only sample', and infer $\eta$ on the unlabeled sample $\{n_i\}_{i=1}^k$.
When inferring $\eta$, however, these methods do not account for the uncertainties on the estimate of $f_b$, which unavoidably impacts the validity of the resulting inference.  

The approach proposed by \citet{banerjee_algeri_unbinned}, outlined in Section \ref{sec:main_argument}, offers a change of perspective that overcomes the need to estimate $f_b$ and associated challenges. Their inferential constructs, however, are limited to the case in which the data available are independent and identically distributed. Sections \ref{sec:YESbkg}-\ref{sec:NObkg} demonstrate how the same approach can be pursued in the context of a binned data analysis.  

\section{Compensator-based inference under model misspecification}
\label{sec:main_argument}
Let $g$ be a postulated background density with distribution function $G$, sharing the same support as $F_s$ and $F_b$, and let $\frac{f_s}{g}, \frac{f_b}{g} \in \Hl$.  To ease the intuition, begin with the case in which $g$ is fully specified. Let $\mathcal{T}=\{1,S^\dag,T_1,T_2,\dots\}$ be an orthonormal basis in $\Hl$ with
\begin{equation}
\label{eqn:basis}
\begin{split}
\quad S^\dag=\frac{S}{\|S\|_{G}},\quad S = \frac{f_s}{g} - 1,
\end{split}
\end{equation}
and the remaining functions $\{T_j\}_{j\geq 1}$ are constructed by orthonormalizing elements of any orthonormal basis in $\Hl$ with respect to $S^\dag$ via Gram-Schmidt process.
Express $f_b$ as:
\begin{equation}
    \label{eq:fb_by_gb}
   f_b(x)=g(x)\biggl[ 1 + \sum\limits_{j=1}^\infty \zeta_j T_j(x) + \delta S^\dag(x)\biggl];
\end{equation}
where the term in the square brackets is an orthonormal expansion of $\frac{f_b}{g}$, and
\[
\delta = E_{F_b}[S^\dag(X)],\quad \zeta_j = E_{F_b}[T_j(X)],\quad\text{for all $j\geq 1$,}
\]
with $E_{F_b}[\cdot]$ being the expectation taken under $F_b$.
The function $S^\dag$ is the normalized score function for   $f$ in \eqref{eqn:true_model} evaluated at $\eta = 0$, when $f_b \equiv g$. It describes the direction from which deviations of the postulated background toward the signal model occur.  By ensuring its inclusion into the basis $\mathcal{T}$,  it allows us to explicitly oversee how such deviations affect the inference on $\eta$. 
In particular, by plugging in \eqref{eq:fb_by_gb} into \eqref{eqn:true_model}, we obtain:
\begin{equation}
    \label{eq:f_by_gb}
 \begin{aligned}
    f(x) = g(x)&\biggl[
    1 + \sum\limits_{j=1}^\infty \tau_j T_j(x) + \theta S^\dag(x)
    \biggl]\quad \text{with}\\
 \theta = E_F[S^\dag(X)]=\eta ||S||_G+&(1-\eta)\delta;
 \quad  \tau_j=E_F[T_j(X)] = (1-\eta)\zeta_j
    \end{aligned}
\end{equation}
for all $j\geq 1$. Hence,
\begin{equation}
    \label{eq:eta_def}
    \eta = \frac{\theta - \delta}{\|S\|_{G} - \delta}.
\end{equation}
In \eqref{eq:f_by_gb}, all $\tau_j$ parameters and $\theta$  are identifiable and from \eqref{eq:eta_def} it follows that, when $\delta=0$, the process of inferring $\eta$ is equivalent to that of inferring $\theta$. When $\delta \ne 0$, $\theta$ can still be employed to infer $\eta$. The parameter $\delta$, however, now intervenes by compensating for the differences between $g$ and $f_b$. For this reason, we refer to such a parameter as `the compensator'. 
While $\delta$ is not identifiable in \eqref{eqn:true_model}, it is under \eqref{eq:fb_by_gb}. Hence, as described in the next section, when a labelled background-only sample is available, asymptotically Gaussian, consistent estimators for $\eta$ can be derived.

\section{Inference when a labeled dataset is available}
\label{sec:YESbkg}
Denote with $\{m_i\}_{i=1}^k$ a collection of (labeled) independent background-only Poisson counts such that $M=\sum_{i=1}^km_i$, $T_b=E_{F_b}[M]$, and    
$$E_{F_b}[m_i]=T_b b_i\quad\text{with}\quad b_i=\int_{\Delta_i}f_b(x)dx.$$
Similarly to \eqref{eqn:sparsness}, we assume:
\begin{equation}
\label{eqn:sparsness2}
\frac{T_b}{k}\rightarrow c_b\in(0,\infty) \quad\text{as} \quad T_b,k\rightarrow \infty,
\end{equation}
that is, also for the background-only samples, the counts are not sufficiently large to be considered approximately Gaussian. From \eqref{eqn:sparsness} and \eqref{eqn:sparsness2} we have:
\begin{equation}
\label{eqn:pi}
\lim_{N, M\rightarrow\infty}\frac{N}{N+M}=\frac{c}{c+c_b} = \pi \in(0,1).\end{equation} Such a limit implies that, as is often realistic in practice, the labeled background-only dataset $\{m_i\}_{i=1}^k$ is large, but not remarkably larger than $\{n_i\}_{i=1}^k$; thus, the uncertainties of estimates based on the former can not be neglected.

Consider the estimators of $\theta$ and $\delta$ given by 
\[
\widehat\theta=\frac{1}{N}\sum_{i=1}^kn_i S^\dag(x_i)\quad \text{and}\quad \widehat\delta=\frac{1}{M}\sum_{i=1}^k m_i S^\dag(x_i);
\]
 Their limiting distribution can be derived using the following lemma. Its proof is given in Appendix \ref{app:proof_lemma1}.
\begin{lemma}\label{prop:lemma1}
Let $\{\xi_i\}_{i=1}^k$ be a deterministic sequence such that
$|\xi_i|\leq C$,  for some $C\in \Real$, for all $i=1,\dots,k$ and, under \eqref{eqn:sparsness}, 
$\sum_{i=1}^kf_i\xi^2_i\rightarrow \sigma^2<\infty$ and  $\sqrt{T}\sum_{i=1}^kf_i\xi_i\rightarrow 0.$
Then, $\frac{1}{\sqrt{N}}\sum_{i=1}^k n_i\xi_i\xrightarrow{d}\N(0,\sigma^2)$.
\end{lemma}
Choosing $\xi_i=\Sd(x_i)-\theta$ or $\xi_i = \Sd(x_i)-\del$ , it is easy to show that the conditions of Lemma \ref{prop:lemma1} hold  when considering either  $\{n_i\}_{i=1}^k$ or $\{m_i\}_{i=1}^k$, under \eqref{eqn:sparsness} and \eqref{eqn:sparsness2}, respectively (see Appendix \ref{app:proof_prop1}). Thus, 
\[
\sqrt{N}(\widehat \theta -\theta)\xrightarrow{d}\N(0,\sigma^2_\theta)\quad\text{and}\quad  \sqrt{M}(\widehat \delta -\delta)\xrightarrow{d}\mathcal{N}(0,\sigma^2_\delta)
\]
with $\sigma^2_\theta=\int(S^\dag(x)-\theta)^2dF(x)$ and   $\sigma^2_\delta=\int(S^\dag(x)-\delta)^2dF_b(x)$.  Inference on $\eta$ can then be performed on the basis of the following proposition.
\begin{proposition}\label{prop:etahat_distr}
Let $\widehat \eta =\frac{\widehat\theta-\widehat\delta }{||S||_G-\widehat\delta}$. If $f_s$ and $g$ are continuously differentiable on $\X$, then under \eqref{eqn:sparsness}, \eqref{eqn:sparsness2} and \eqref{eqn:pi}:
\begin{equation}
\label{eqn:etahat_distr}
\mathcal{Z}_1=\frac{\sqrt{MN}(\widehat\eta-\eta)}{\sqrt{\frac{M}{(||S||_G-\widehat\delta)^2}\widehat\sigma^2_\theta+\frac{N(||S||_G-\widehat\theta)^2}{(||S||_G-\widehat\delta)^4}\widehat\sigma^2_\delta}}\dconv\mathcal{N}(0,1)
\end{equation}
with $\widehat\sigma^2_\theta=\frac{1}{N}\sum_{i=1}^kn_i(S^\dag(x_i))^2-\widehat\theta^2$ and $\widehat\sigma^2_\delta=\frac{1}{M}\sum_{i=1}^km_i(S^\dag(x_i))^2-\widehat\delta^2$.
\end{proposition}
The proof is given in Appendix \ref{app:proof_prop1}. Proposition \ref{prop:etahat_distr} implies that an asymptotic $p$-value to test \eqref{eqn:eta_test} corresponds to the right tail probability of a standard normal evaluated at the value of $\mathcal{Z}_1$ observed on the data at hand with $\eta=0$. 

As demonstrated in Section \ref{sec:fermi_YESbkg} with an example, when $\eta$ is inferred via the test statistic $\mathcal{Z}_1$ in \eqref{eqn:etahat_distr}, the impact of the choice of $G$ becomes negligible. That is because the departures of the latter from $F_b$ are fully accounted for through the estimation of $\delta$. 
Nonetheless, it is possible to extend Proposition \ref{prop:etahat_distr} to allow the postulated background distribution to depend on an unknown parameter $\beta\in \mB \subset \mathbb{R}^p$. 
To emphasize the dependence on the latter, we write $G_\beta$, and we let $g_{\beta}$ be the corresponding density. Since the function $S$, as defined in \eqref{eqn:basis} and all related quantities, also depend on $\beta$ through $g_{\beta}$, we denote with $S_{\beta}$, $\|S_\beta\|_{G_{\beta}}^2$, $S^\dag_\beta$, $\theta_\beta$, and $\delta_\beta$ the parametric  counterparts of $S$, $\|S\|_{G}^2$, $S^\dag$, $\theta$, and $\delta$, respectively.

In the remainder of this section, we shall assume that the regularity conditions \ref{assump:gb_concave}-\ref{assump:fs_cont_diff}, listed in Appendix \ref{app:assumptions} and \ref{app:proof_prop2} hold. In the Poisson binned data regime,  such conditions are inherently equivalent to the standard regularity conditions used in the asymptotic analysis of parametric families of distributions \citep[cf.][Ch. 5]{van2000asymptotic} in the i.i.d. setting. 

Denote with $\widehat{\beta}$ the maximum likelihood estimator of $\beta$, evaluated on the background-only sample, i.e., 
\begin{equation}
\label{eqn:MLE}\widehat\beta=\arg\max_{\beta}\sum_{i=1}^km_i\log g_{\beta i}\end{equation}
with $g_{\beta i}=\int_{\Delta_i}g_\beta(x)dx$. Let $\beta^*$ be the minimizer of the Kullback - Leibler divergence between $F_b$ and $G_\beta$. As proven in Appendix \ref{app:proof_bt_conv} of the $\bth \pconv \bts$. The parameters $\theta_{\bts}$ and $\delta_{\bts}$ can then be estimated via
\[
\thth_{\bth} = \frac{1}{N} \sumk n_i \Sd_{\bth}(x_i) \quad \text{and} \quad
    \delh_{\bth} = \frac{1}{M} \sumk m_i \Sd_{\bth}(x_i),
\]
respectively, and an estimator of $\eta$ is:
\[
\widehat \eta_{\bth} =\frac{\thth_{\bth} - \delh_{\bth}}{||S_{\bth}||_{G_{\bth}}-\thth_{\bth}}.
\]
Proposition \ref{prop:etahatb_distr} describes the asymptotic distribution of $\widehat\eta_{\bth}$ when its limiting variance is estimated using a consistent estimator, $\sigh^2_{\bth,\widehat{\eta}}$. The proof of the proposition and the expression for $\sigh_{\bth,\widehat{\eta}}$ are given in Appendix \ref{app:proof_prop2}.

\begin{proposition}
\label{prop:etahatb_distr}
Under \eqref{eqn:sparsness}, \eqref{eqn:sparsness2} and \eqref{eqn:pi},
\begin{equation}
\label{eqn:etahatb_distr}
\mZ_2 = \sqrt{\frac{MN}{M+N}}\frac{(\widehat\eta_{\bth}-\eta)}{\sigh_{\bth, \eta}}\dconv\mathcal{N}\big(0,1\big).
\end{equation}
\end{proposition}
An asymptotic $p$-value for testing \eqref{eqn:eta_test} is given by the right tail probability of a standard normal distribution, evaluated at the observed value of $\mZ_2$ with $\eta = 0$. Different from $\mZ_1$  in \eqref{eqn:etahat_distr}, the test statistic $\mZ_2$ accounts for the uncertainties associated with the estimation of $\bt$.

\subsection{Case study: signal detection at the Fermi Large Area Telescope}
\label{sec:fermi_YESbkg}

We consider a realistic simulation of data generated by the Fermi Large Area Telescope (LAT), also analyzed in \citep{banerjee_algeri_unbinned}. The Fermi LAT is a $\gamma$-ray telescope on the orbiting Fermi satellite \citep{atwood2009large}. 
The data under study consist of two energy spectra: one corresponding to the situation in which only the astrophysical background is present, and the other in which a dark matter signal has been injected on top of it. The former has size $M=4427$, serves as a background-only sample, and is generated from a Pareto type I distribution with shape 1.4 and truncated over the interval $[1,35]$ GeV. The latter constitutes the physics sample and has size $N=2338$. The signal of interest is modeled as a Gaussian bump, centered at $3.5$GeV with standard deviation $0.35$GeV. Both samples include representations of detector effects and systematic errors. 

Given the detector's high resolution, the original dataset could be treated as unbinned \citep[see][]{banerjee_algeri_unbinned}; however, to assess how the resolution of different binning schemes may affect the inferential results, here we choose to discretize the original spectrum into $k = 30, 50$, and 100 bins. Moreover, for visualization purposes, we analyze the data in log-scale. 

For the log-transformed data,  $f_b$ and $f_s$ in \eqref{eqn:true_model} specify as:

\[
 f_b(x) \propto \exp(-1.4x)  \quad \text{and}\quad f_s(x) \propto \exp\Big\{-\frac{(\exp(x)-3.5)^2}{2(0.35)^2}\Big\} \exp(x), 
\]
with $x \in [0,\log(35)]$. The tests based on \eqref{eqn:etahat_distr} and \eqref{eqn:etahatb_distr} are then performed using four different choices for the proposal background that consists of two fully specified densities and two parametric forms of $g_\bt$. The former two choices consist of a uniform distribution on the interval $[0,\log(35)]$ and a truncated exponential distribution with the rate parameter fixed at 0.5. The parametric choices for $g_\bt$ include a truncated exponential but with unknown rate and a truncated Gaussian density with mean -1 and unknown variance.

For these two parametric distributions, the unknown parameters are estimated via MLE on the background-only data. As shown in Figure \ref{fig:Fermi_LAT_g} (left panel), the four proposal backgrounds considered exhibit different degrees of deviation from $f_b$. Naturally, the exponential density with estimated rate (blue long-dashed line) provides the closest approximation to $f_b$ (black solid line), while the Gaussian tail (brown dashed line) exhibits a slower decay. On the contrary, the exponential density with fixed rate (purple dot-dashed line) and the uniform density (red two-dashed line) differ substantially from the truth.

\begin{table}[t]
    \centering
\renewcommand{\arraystretch}{1.25}
 \begin{tabular}{|c|c|c|c|}
\hline
     Bins & Postulated background & $\mathbf{\widehat\eta}$ & $p$\textbf{-value}\\ \hline \hline
     % ######################### k = 30 #################################
     \multirow{3}{*}{ $k = 30$} &
     Exp($\bth$) & 0.0417 & $1.3410 \times 10^{-6}$ \\ \cline{2-4} &
     Gaussian-tail($\bth$) & 0.0417 & $1.3666 \times 10^{-6}$ \\ \cline{2-4} &
     Exp(0.5) & 0.0424 & $1.0377 \times 10^{-6}$ \\ \cline{2-4} &
     Uniform & 0.0428 & $9.3480 \times 10^{-7}$ \\
     \hline \hline
     % ######################### k = 50 #################################
     \multirow{3}{*}{ $k = 50$} &
     Exp($\bth$) & 0.0419 & $1.1238 \times 10^{-6}$ \\ \cline{2-4} &
     Gaussian-tail($\bth$) & 0.0419 & $1.1473 \times 10^{-6}$ \\ \cline{2-4} &
     Exp(0.5) & 0.0426 & $8.0030 \times 10^{-7}$ \\ \cline{2-4} &
     Uniform  & 0.0430 & $6.8600 \times 10^{-7}$ \\
     \hline \hline
     % ######################### k = 100 #################################
     \multirow{3}{*}{ $k = 100$} &
     Exp($\widehat\beta$) & 0.0425 & $8.761 \times 10^{-7}$ \\ \cline{2-4} &
     Gaussian-tail($\bth$) & 0.0425 & $8.870 \times 10^{-7}$ \\ \cline{2-4} &
     Exp(0.5) & 0.0429 & $7.348 \times 10^{-7}$ \\ \cline{2-4} &
     Uniform & 0.0431 & $6.926 \times 10^{-7}$ \\
     \hline
\end{tabular}
    \caption{Comparing the estimates of $\eta$ and the $p$-values of the tests based on \eqref{eqn:etahat_distr} and \eqref{eqn:etahatb_distr} obtained on the Fermi LAT data in the presence of a background-only sample for different proposal background distributions with $k = 30, 50$ and 100 bins.}
    \label{tab:fermi_lat_wbkg}
\end{table}

Table \ref{tab:fermi_lat_wbkg} reports the results of the test conducted using the four different choices of the proposal background considered with different binning resolutions.
For all values of $k$ considered, different choices of the proposal background yield similar estimates of $\eta$ with small $p$-values ($O(10^{-6})$ or $O(10^{-7})$). More importantly, the estimated signal intensity and the $p$-values obtained when choosing an exponential background are rather similar regardless of whether the rate is estimated or fixed to a value far from the truth.  This suggests that, while the use of a parametric postulated background density is possible, it is somewhat unnecessary as long as the compensator $\delta$ is efficiently estimated. Lastly, the uniform distribution, despite exhibiting the most deviation in shape from the true background, also produces an estimate of $\eta$ consistent with all other cases and similar $p$-values. This further demonstrates the robustness of the inferential framework with respect to the choice of the proposal background distribution. The signal search outcome remains consistent as the number of bins varies from 30 to 100, with the case $k = 100$ showing the most resemblance with the unbinned analysis presented in \citealt{banerjee_algeri_unbinned}.

\section{Bump hunting without background-only data}
\label{sec:NObkg}
When an unlabeled, background-only sample is not available, the compensator is no longer estimable. Nonetheless, \eqref{eq:eta_def} implies that:

\[
\theta_{0,\beta} = \frac{\theta_{\beta}}{\|S_{\beta}\|_{G_{\beta}}} \le \eta\quad \text{when}\quad \delta_\beta\leq 0.
\]

\begin{center}
    \includegraphics[width=0.45\linewidth]{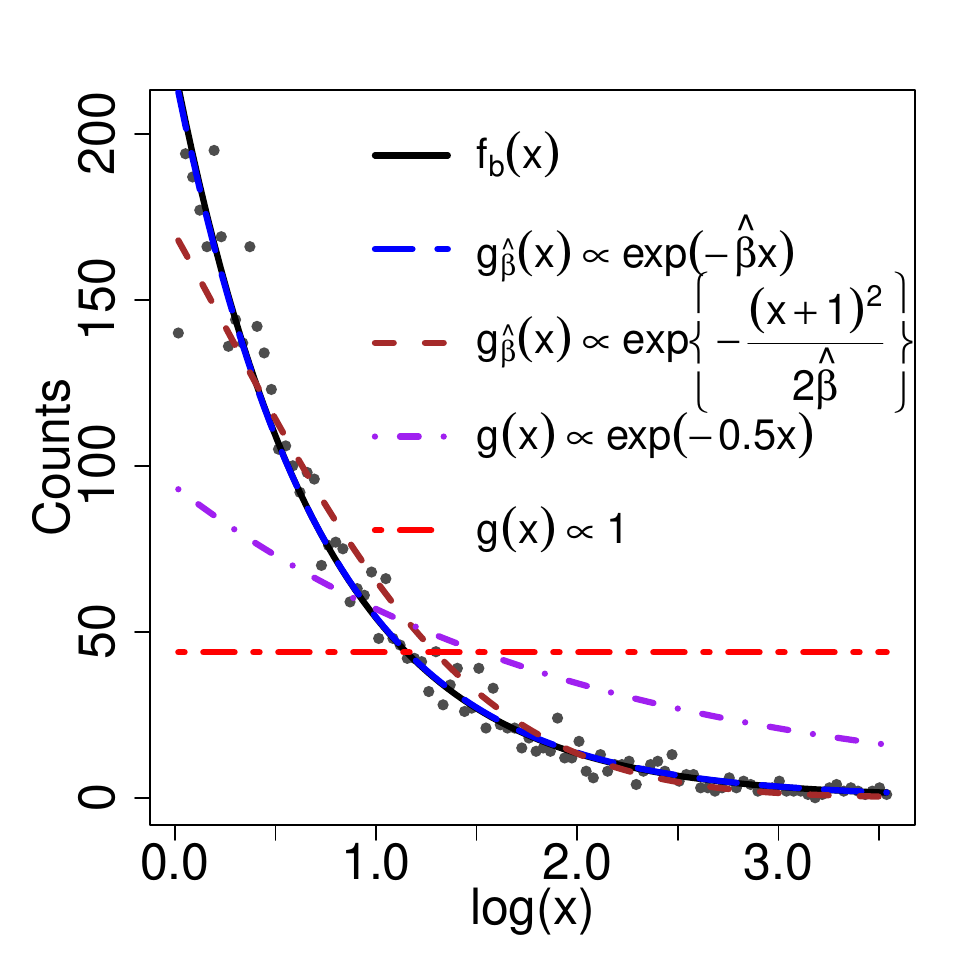}\hspace{-0.5cm}
    \includegraphics[width=0.45\linewidth]{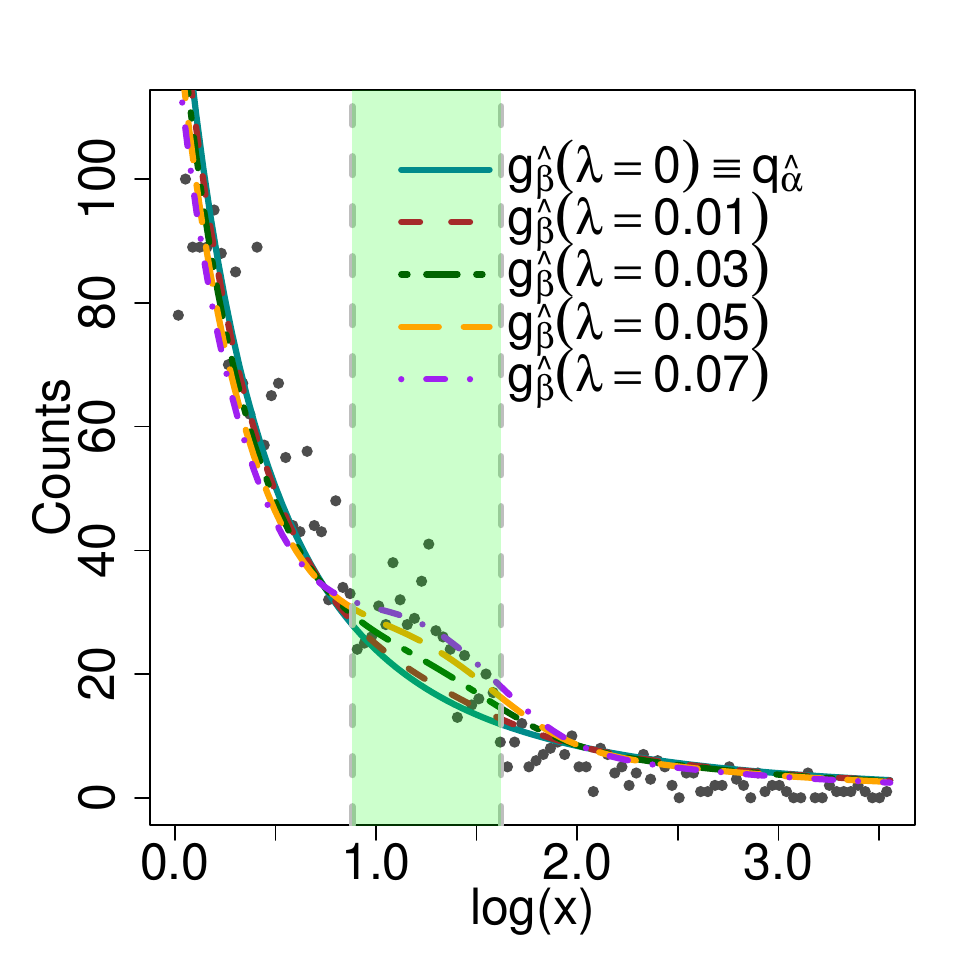}
    \captionof{figure}{Left panel: Graphs of four different choices of the postulated background density: a uniform density (red two-dashed line), an exponential density with rate fixed at 0.5 (purple dot-dashed line), an exponential density with unknown rate parameter (blue long-dashed line), and a Gaussian density with mean -1 and unknown variance (brown dashed line), all truncated on the interval $[0, \log(35)]$. In the last two cases, the unknown parameters are estimated via MLE. The densities have been appropriately scaled and overlaid on the plot showing bin counts (considering $k= 100$ bins) from the background-only sample transformed into log-scale. Right panel: example of the proposed sensitivity analysis conducted on the Fermi LAT data.}
    \label{fig:Fermi_LAT_g}
\end{center}

Hence, inferring $\theta_{0,\beta}$ in place of $\eta$ 
yields conservative, yet valid, inference on the latter.  

\citet{banerjee_algeri_unbinned} identify sufficient conditions to ensure non-positivity of $\delta_\beta$
for the case of one-dimensional bump-hunting problems, that is, when searching for a localized bump on top of a smoothly decaying background. This is the case, for example, in the Fermi LAT analysis proposed in Section \ref{sec:fermi_YESbkg} and in several other problems encountered in the physical sciences and beyond \citep[e.g.,][]{dvd, collins2019extending, volkovich2022data}.  

Let $M_\epsilon = [\mu_s - d_\epsilon, \mu_s + d_\epsilon]\subset\Real$ be the subset of $\mathcal{X}$ over which $F_s$ concentrates its mass so that $F_s(M_\epsilon) = 1-\epsilon$ for some $\epsilon>0$. For example, in our Fermi LAT example, this corresponds to the green shaded region in Figure \ref{fig:Fermi_LAT_g} (right panel). \citet{banerjee_algeri_unbinned} show that   $\delta_\beta\leq 0$ whenever 
\[
\sup_{x \not\in M_\epsilon} \frac{f_b(x)}{g_\beta(x)} =o(\epsilon^{-1})\quad \text{and} \quad \sup_{x \in M_\epsilon} \frac{f_b(x)}{g_\beta(x)} \le 1.
\]
Hence, the goal is to identify a postulated background density, $g_\beta$, that is (i) `not too far' from $f_b$ outside $M_\epsilon$, so that, for sufficiently small $\epsilon$, the first of the above conditions is verified, and, at the same time, (ii) bounds $f_b$ for all $x\in M_\epsilon$. Since $f_b$ is unknown, a sensitivity analysis to assess the validity of (i)-(ii) can be performed for suitably constructed choices of  $g_\beta$. 
One possible choice, among many, is described below.

Let $\beta= (\alpha, \lambda)$, with $\alpha\in \mathcal{A}\subseteq \Real^{p-1}$ characterizing the shape of a smoothly decaying density, $q_\alpha$, that serves as a baseline model to approximate $f_b$; whereas, $\lambda$ controls the size of a dominating term introduced in $g_\beta$ to bound $f_b$ over $M_\epsilon$. In particular:
\begin{equation}
    \label{eqn:g_lambda}
    g_\beta(x) = (1-2\lambda) q_\alpha(x) + \lambda\Big[
    \phi_{\mathcal{X}}(x; \mu_1, \sigma_0) + \phi_{\mathcal{X}}(x; \mu_2, \sigma_0)
    \Big];
\end{equation}
with $\beta = (\alpha, \lambda) \in \mathcal{A} \times [0,1/2)$,  and $\phi_{\mathcal{X}}(\cdot; \mu, \sigma)$ denotes the Gaussian density with mean $\mu$ and standard deviation $\sigma$, truncated on $\mathcal{X}$. To ensure $g_\beta$ bounds $f_b$ over $M_\epsilon$,  $\mu_1$ and $\mu_2$ are chosen sufficiently close to the boundaries of such a region; whereas, $\sigma_0$ should be substantially larger than the width of the true signal and described by $f_s$. 
Once $\alpha$ is replaced by its maximum likelihood estimate, $\widehat{\alpha}$,  obtained by fitting $q_\alpha$ on the physics data at hand, a sensitivity analysis is performed to visually assess which values of $\lambda$ yield  \eqref{eqn:g_lambda} to dominate $f_b$ over $\mathcal{X}$.

For example, when considering the Fermi LAT data described in the previous section, such a sensitivity analysis can be performed on the basis of the right panel of Figure \ref{fig:Fermi_LAT_g} in which a diffused dominating term is injected, using different values of $\lambda$ in \eqref{eqn:g_lambda}, on top of the estimated baseline density  $q_{\alh}$ (solid cyan line). Details are given in Section \ref{sec_Fermi_nobkg}. 

Once a suitable value of  $\lambda$, namely $\lambda^*$,  is selected through the sensitivity analysis, we test:
\begin{equation}
    \label{eqn:tht_0bs_test}
    H_{0,\theta_0}: \theta_{0,\beta^*} = 0 \quad \text{vs} \quad
    H_{1,\theta_0}:\theta_{0,\beta^*}  >0,
\end{equation}
in which $\beta^* = (\alpha^*, \lambda^*)$ with $\alpha^*$ denoting the asymptotic limit of $\widehat\alpha$ under $F$. For what follows, we assume that the regularity conditions \ref{assump:gb_concave}-\ref{assump:fs_cont_diff}, described in the Appendix, are satisfied by the baseline model $\{q_{\alpha}:\alpha \in \mathcal{A}\}$. Under such conditions, $\alpha^*$ is the minimizer of the Kullback-Leibler divergence between $F$ and the CDF of $q_\alpha$.

A natural estimator for $\tht_{0,\bts}$ is  
\[
\thth_{0,\bth} = \frac{\thth_{\bth}}{\|S_{\bth}\|_{G_{\bth}}}
\]
with $\bth = (\alh, \lambda^*)$.
Let $\sigh_{\bth,\thth_{0}}$ be a consistent estimator of the limiting variance of $\thth_{0,\bth}$, the asymptotic distribution of the latter is given by Proposition \ref{prop:thth_0bts_distr}. Its proof and an explicit expression for $\sigh_{\bth,\thth_{0}}$ are given in Appendix \ref{app:proof_prop3} of the.
\begin{proposition}
\label{prop:thth_0bts_distr}
    Under \eqref{eqn:sparsness}, 
    \begin{equation}
        \label{eqn:thth_bth_dist}
        \mZ_3 = \frac{\sqrt{N}\Big(\thth_{0,\bth} - \tht_{0,\bts}\Big)}{\sigh_{\bth,\thth_0}} \dconv
        \N(0, 1).
    \end{equation}
\end{proposition}
The test in \eqref{eqn:tht_0bs_test} can be conducted by setting  $\theta_{0,\bts} = 0$  in \eqref{eqn:thth_bth_dist} and the corresponding asymptotic $p$-value is again calculated as the right tail probability under a standard normal distribution from the observed value of $\mZ_3$. 

\begin{table}[!t]
    \centering
 \begin{tabular}{|c|c|c|c|}
\hline
     Bins & $\lambda$ & $\mathbf{\widehat\tht_{0,\bth}}$ & $p$\textbf{-value}\\ \hline \hline
     % ######################### k = 30 #################################
     \multirow{3}{*}{ $k = 30$} 
     & 0.03 & 0.0359 & $7.39 \times 10^{-7}$\\ \cline{2-4}
     & 0.05 & 0.0232 & $1.09 \times 10^{-3}$\\ \cline{2-4}
     & 0.07 & 0.0100 & $9.53 \times 10^{-2}$\\
     \hline \hline
     % ######################### k = 50 #################################
     \multirow{3}{*}{ $k = 50$} 
     & 0.03 & 0.0352 & $1.08 \times 10^{-6}$\\ \cline{2-4}
     & 0.05 & 0.0225 & $1.43 \times 10^{-3}$\\ \cline{2-4}
     & 0.07 & 0.0093 & $1.11 \times 10^{-1}$\\
     \hline \hline
     % ######################### k = 100 #################################
     \multirow{3}{*}{ $k = 100$}
     & 0.03 & 0.0359 & $7.11 \times 10^{-7}$\\ \cline{2-4}
     & 0.05 & 0.0232 & $1.07 \times 10^{-3}$\\ \cline{2-4}
     & 0.07 & 0.0100 & $9.50 \times 10^{-2}$\\ 
     \hline
\end{tabular}
    \caption{
    Results obtained in the Fermi LAT data for the test based on \eqref{eqn:thth_bth_dist} choosing different values of $\lambda$ and different binning resolutions.
    } 
    \label{tab:fermi_lat_wobkg}
\end{table}

\subsection{Case study: Analyzing Fermi LAT data without background-only data}
\label{sec_Fermi_nobkg}
Let us consider the Fermi LAT data introduced in Section \ref{sec:fermi_YESbkg}, but let us now assume that only the physics sample is available. Choose $g_\bt$ as in \eqref{eqn:g_lambda} with $q_{\al}$ specifying as

\[
q_{\al}(x) \propto (x+1)^{-(\al+1)}, \quad 0 \le x \le \log(35),
\]
$\epsilon = 0.001$, $\mu_1 = 1.067$, $\mu_2 = 1.437$, and $\sigma_0 = 0.304$ (three times the standard deviation of $f_s$). The parameter $\al$ is estimated on the physics data via MLE, yielding $\alh = 1.588$; the corresponding curve is displayed as a cyan solid line on the right panel of Figure \ref{fig:Fermi_LAT_g}. The density $g_\bt$ obtained from choosing different values of $\lambda$ is also displayed on the same plot to enable the desired sensitivity analysis. In particular, the plot suggests that $g_{\bth}$ begins to exhibit a prominent dominating component at $\lambda = 0.03$ (dark green two-dashed line). Therefore, it is sensible to assume that, for  $\bth = (\alh, 0.03)$, $g_{\bth}$ dominates $f_b$ over $M_\epsilon$ (green shaded area), leading to a negative value for $\delta_{\bts}$. 

Table \ref{tab:fermi_lat_wobkg} reports the estimates obtained for  $\theta_{0,\beta*}$ and the $p$-values of the test based on \eqref{eqn:thth_bth_dist} when $\lambda = 0.03, 0.05, 0.07$ and with varying number of bins ($k = 30, 50, 100$). 
For all three binning resolutions, the results are similar. 
Focusing on the case in which  $k = 100$, when $\lambda$ is chosen to be $0.03$, the test based on \eqref{eqn:thth_bth_dist} yields a signal discovery with estimated intensity $\thth_{0,\bth} = 0.0359$ and a $p$-value of $7.11 \times 10^{-7}$. Alternatively, in order to ascertain the negativity of the compensator, one may choose $\lambda = 0.05$ (orange long-dashed line). In this case, $g_{\bth}$ shows a more prominent dominating component, and the estimated signal intensity is, as expected,  more conservative ($\thth_{0,\bth} = 0.0232$). While this test exhibits lower significance, the $p$-value is still sufficiently small ($1.07 \times 10^{-3}$) to provide statistical evidence in support of the signal. On the other hand, $\lambda = 0.07$ (purple dot-dashed line) appears to yield an excessively large dominating component and, unsurprisingly, the corresponding $p$-value is too large to claim a discovery (0.095). 

\section*{Supplementary material and data availability}
The \texttt{R} code and the data used for the analysis presented in this manuscript are openly available at 
\newline 
\href{https://github.com/baner175/binned_signal_detection}{https://github.com/baner175/binned\_signal\_detection}.

\section*{Acknowledgments}
The authors are grateful to Oliver Rieger and Lydia Brenner for the valuable feedback and discussions. The interactions with them motivated us to extend \citep{banerjee_algeri_unbinned} to the binned data setting.  

AB's work has been partially funded by the Errett W. McDiarmid Graduate Fellowship, provided by the College of Liberal Arts at the University of Minnesota. 
SA's work has been partially funded by the Warwick MidCareer Faculty Research Award, College of Liberal Arts, University of Minnesota. 

\appendix
\section{Proof of main results}
\label{app:technical_proofs}

\subsection{Proof of Lemma \ref{prop:lemma1}}
\label{app:proof_lemma1}
For some $t\in \R$, the moment generating function of $\frac{1}{\sqrt{T}}\sumk n_i\xi_i$ is:
\begin{equation}
    \label{eqn:MGF_expansion}
    \begin{aligned}
        &\mE\Bigg[\exp\Big(
        \frac{t}{\sqrt{T}}\sumk n_i\xi_i
        \Big)\Bigg] = \prod_{i=1}^k \mE\Bigg[\exp\Big(\frac{t\xi_i}{\sqrt{T}}n_i\Big)\Bigg] 
        \\
        &=
        \prod_{i=1}^k \exp\Bigg[Tf_i\Bigg\{\exp\Big(\frac{t\xi_i}{\sqrt{T}}\Big)-1\Bigg\}\Bigg] \\
        &=
        \prod_{i=1}^k \exp\Bigg[Tf_i\sum_{j\ge 1}\frac{T^{-j/2}(t\xi_i)^j}{j!}\Bigg] \\
        &=
        \exp\Bigg[
        t\sqrt{T}\sumk f_i\xi_i + \frac{t^2}{2}\sumk \xi_i^2f_i + 
        \sumk \sum_{j \ge 3} \frac{T^{1-j/2}t^j\xi_i^j}{j!}f_i
        \Bigg],
    \end{aligned}
\end{equation}
where the last term in the exponent converges absolutely to zero under \eqref{eqn:sparsness}. Specifically,
\[
\sum_{j \ge 3} \sumk \Bigg| \frac{T^{1-j/2}t^j\xi_i^j}{j!}f_i\Bigg| \le
\frac{1}{\sqrt{T}}\sum_{j \ge 3} \frac{C^j|t|^j}{j!}\sumk f_i < \frac{e^{C|t|}}{\sqrt{T}} \to 0 \quad \text{as $T \to \infty$}.
\]
Therefore, 
\[
\mE\Bigg[\exp\Big(
        \frac{t}{\sqrt{T}}\sumk n_i\xi_i
        \Big)\Bigg] \to \exp\Big(\frac{\sig^2t^2}{2}\Big),
\]
implying that $\frac{t}{\sqrt{T}}\sumk n_i\xi_i \dconv \mathcal{N}(0,\sigma^2)$. Since, $N/T \pconv 1$, the statement of the lemma follows by Slutsky's theorem. 

\hfill $\square$

\subsection{Proof of Proposition \ref{prop:etahat_distr}}
\label{app:proof_prop1}
Set $\nu = (\tht_0, \del_0)$ with
\begin{equation}
    \label{eqn:tht0_del0_def}
    \tht_0 = \frac{\tht}{\|S\|_{G}} = \intmx S_0(x)dF(x), \quad
    \del_0 = \frac{\del}{\|S\|_{G}} = \intmx S_0(x)dF_b(x),
\end{equation}
and $S_0(x) = \Sd(x)/\|S\|_G$. The corresponding estimator is $\nuh= (\thth_0, \delh_0)$ with
\begin{equation}
    \thth_0 = \frac{\thth}{\|S\|_{G}} = \frac{1}{N} \sumk n_iS_0(x_i) \quad \text{and}\quad
    \delh_0 = \frac{\delh}{\|S\|_{G}} = \frac{1}{M}\sumk m_i S_0(x_i).
\end{equation}
Therefore,
\[
\sqrt{N}(\thth_0 - \tht_0) = \frac{1}{\sqrt{N}}\sumk \xi_i, \text{ with } \quad  \xi_i = S_0(x_i) - \tht_0.
\]
Since $f_s$ and $g$ are continuously differentiable, $S_0$ and the coordinates of the gradient vector function $\partial S_0 = (\partial_1S_0, \partial_2S_0, \cdots, \partial_dS_0)$ are also continuous on $\X$ and thus bounded, leading to the $\xi_i$'s being bounded as well. 

Under \eqref{eqn:sparsness}, the Cauchy-Schwartz inequality yields
\begin{equation}
    \label{eqn:fi_xi_conv}
    \begin{aligned}
    \Big|\sqrt{T}\sumk f_i\xi_i\Big| =&
    \sqrt{T}\Big|\sumk \int_{\Delta_i}\Big(S_0(x_i) - S_0(x)\Big)f(x)dx\Big| \\
    =&
    \sqrt{T}\Big|\sumk \int_{\Delta_i}\partial S_0(x_i')^T(x - x_i')f(x)dx\Big|
    \\
    \le&
    \|\X\|_d\sup_{\substack{x \in \X \\ 1 \le i \le d}} \big|\partial_i S_0(x)\big|\frac{\sqrt{Td}}{k} \to 0
    \end{aligned}
\end{equation}
where each $x_i'$ is a point lying between $x_i$ and $x$. Moreover,
\begin{equation}
    \begin{aligned}
    \label{eqn:fi_xi2_conv}
        \sumk f_i \xi_i^2 =& \sumk \Big(S_0(x_i) - \tht_0\Big)^2\int_{\Delta_i} f(x)dx\\
        =& \sumk \big(S_0(x_i) - \tht_0\big)^2 f(x_i)v + o(1)
        = \sig_{0,\tht}^2 + o(1)
    \end{aligned}
\end{equation}
where $\sig_{0,\tht}^2 = \intmx \big(S_0(x_i) - \tht_0\big)^2 dF(x) = \sig_{\tht}^2/\|S\|_G^2$. From Lemma \ref{prop:lemma1} it follows that
\begin{equation}
    \label{eqn:thth0_dist}
    \sqrt{N}(\thth_0 - \tht_0) \dconv \N(0, \sig_{0,\tht}^2).
\end{equation}
Similarly, one can show that, under \eqref{eqn:sparsness2},
\begin{equation}
    \label{eqn:delh0_dist}
    \sqrt{M}(\delh_0 - \del_0) \dconv \N(0, \sig_{0,\del}^2),
\end{equation}
with $\sig_{0,\del}^2 = \intmx (S_0(x) - \del_0)^2dF_b(x) = \sig_{\del}^2/\|S\|_G^2$.

Now, define
\begin{equation}
\label{eqn:T_def}
    W(x,y) = \frac{x-y}{1-y}
\end{equation}
such that $\eta = W(\nu)$ and denote the gradient vector of $W$ by $\partial W = (\pone W, \ptwo W)$ with
\[
\pone W(x,y) = \frac{1}{1-y} \quad \text{and} \quad 
\ptwo W(x,y) = \frac{x-1}{(1-y)^2}.
\]
By the mean value theorem:
\begin{equation}
    \label{eqn:MVT_T_nu}
    \begin{aligned}
        &\sqrt{\frac{MN}{M+N}}(\widehat\eta - \eta) = \sqrt{\frac{MN}{M+N}}\Big(W(\nuh) - W(\nu)\Big)\\
        =&
        \sqrt{\frac{M}{M+N}}\sqrt{N}(\thth_0 - \tht_0)\pone W(\nut) + 
        \sqrt{\frac{N}{M+N}}\sqrt{M}(\delh_0 - \del_0)\ptwo W(\nut)
    \end{aligned}
\end{equation}
where $\nut$ is a point between $\nu$ and $\nuh$. From \eqref{eqn:thth0_dist} and \eqref{eqn:delh0_dist} we have $\nuh \pconv \nu$, hence, $\nut \pconv \nu$. The function $W$ being continuously differentiable implies that $W(\nut) \pconv W(\nu)$. Since the random variables in \eqref{eqn:thth0_dist}-\eqref{eqn:delh0_dist}  are independent, by Slutsky's theorem we have
\begin{equation}
    \label{eqn:eta_hat_dconv}
    \sqrt{\frac{MN}{M+N}}(\widehat\eta - \eta) \dconv \N(0, \sig_{\eta}^2)
\end{equation}
where
\[
\sig_{\eta}^2 
= \frac{(1-\pi)\sig_{\tht}^2}{(\|S\|_G - \del)^2} + \frac{\pi\sig_{\del}^2(\tht - \|S\|_G)^2}{(\|S\|_G-\del)^4}.
\]
It can be easily shown that $\sigh_{\tht}^2$ and $\sigh_{\del}^2$ are consistent estimators for $\sig_{\tht}^2$ and $\sig_{\del}^2$, respectively. A consistent estimator $\sigh_{\eta}^2$ for $\sig_{\eta}^2$ can then be defined by substituting $\frac{N}{M+N}$, $\sigh_{\tht}^2$, and $\sigh_{\del}^2$ in place of  $\pi$, $\sig_{\tht}^2$, and $\sig_{\del}^2$ in the above expression. Finally, dividing the left-hand side of \eqref{eqn:eta_hat_dconv} by $\sigh_{\eta}$, \eqref{eqn:etahat_distr} follows by Slutsky's theorem.

\hfill $\square$

\subsection{Regularity conditions required in the parametric setting}
\label{app:assumptions}
For the results proven in the sections that follow, we assume:
\begin{enumerate}[label=\textbf{(A\arabic*)}, leftmargin=2em]
    \item For any $x \in \mathcal{X}$, the map $\bt \mapsto \log g_\bt(x)$ is concave. \label{assump:gb_concave}
    \item For any $x \in \mathcal{X}$, the maps $\bt \mapsto g_\bt(x)$, $\bt \mapsto \pb g_{\bt}(x)$, and $\bt \mapsto \pb^2g_\bt(x)$ are continuous in $\bt$, with $\pb$ and $\pb^2$ denoting, respectively, the $p$-dimensional gradient vector, and the $p \times p$ Hessian matrix of a function with respect to the parameter $\bt$.\label{assump:gb_cont_bt}
    \item For any $\bt \in \mathcal{B}$, the maps $x \mapsto g_\bt(x)$, 
    $x \mapsto \pb g_{\bt}(x)$, and $x \mapsto \pb^2 g_{\bt}(x)$ are continuous and differentiable in $x$.\label{assump:gb_cont_x}
    \item The parameter space $\mB \subset \R^p$ is compact.\label{assump:mB_compact}
\end{enumerate}

Observe that, since the search region $\X$ is compact, assumptions \ref{assump:gb_concave}-\ref{assump:mB_compact} imply that $g_{\bt}$ $\pb \log g_{\bt}$, $\pb^2 \log g_{\bt}$, $S_\bt$ etc. are bounded for any fixed $\bt$. This allows us to interchange the derivative with respect to $\bt$ and the integral under $F$, or $F_b$, by the dominated convergence theorem.

\subsection{Proving that $\bth \pconv \bts$}
\label{app:proof_bt_conv}
Since $\pb \log g_{\bt}(x)$ is concave in $\bt$ by \ref{assump:gb_concave}, its MLE, $\bth$, in \eqref{eqn:MLE} and $\bts$ can equivalently be defined as the roots of the estimating equations
\begin{align}
    \label{eqn:bt_score}
    Q_{k}(\bt) = \frac{1}{T_b}\sumk m_i\pb\log g_{\bt i} = 0
\end{align}
and 
\begin{equation}
    Q(\bt) = \mE_{F_b}\Big(\pb\log g_{\bt}(Y)\Big) = 0,
\end{equation}
respectively. Observe that, for any $\bt \in \mB$, under \eqref{eqn:sparsness} and \eqref{eqn:sparsness2},
\begin{equation}
    \label{eqn:EQb_conv}
    \begin{aligned}
        & \mE_{F_b}\Big[Q_{k}(\bt)\Big] =
    \sumk b_i\frac{\pb g_{\bt i}}{g_{\bt i}}
    =
    \sumk \int_{\Delta_i}\pb f_b(x)dx\frac{\int_{\Delta_i}\pb g_{\bt}(x)dx}{\int_{\Delta_i} g_{\bt}(x)dx}\\
    =&
    \sumk v f_b(x_i)\frac{\pb g_{\bt}(x_i)v}{g_\bt(x_i)v} + o(1)
    =
    \intmx \log g_{\bt}(x)dF_b + o(1) = Q(\bt) + o(1),
    \end{aligned}
\end{equation}
and 
\begin{equation}
    \label{eqn:VQb_conv}
\begin{aligned}
     & Var_{F_b}\Bigg[\frac{1}{T_b}\sumk m_i\pb\log g_{\bt i}\Bigg] =
    \frac{1}{T_b}\sumk b_i \Bigg(\frac{\pb g_{\bt i}}{g_{\bt i}}\Bigg)^2\\
    = &
    \frac{1}{T_b} \sumk v f_b(x_i)\Bigg(\frac{\pb g_\bt(x_i)}{g_\bt(x_i)}\Bigg)^2 + o(1) = o(1).
\end{aligned}
\end{equation}
Therefore, $Q_k(\bt) \pconv Q(\bt)$. Furthermore, assumptions \ref{assump:gb_cont_bt}-\ref{assump:mB_compact} ensure $Q(\bt)$ is uniformly continuous and $Q_k(\bt)$ is Lipschitz continuous. Hence, $Q_k(\bt)$ converges uniformly to $Q(\bt)$ in probability \citep[cf.][Corollary 2.2]{newey1991uniform} and the desired convergence follows by Theorem 5.9 in \citeauthor{van1998asymptotic}~(\citeyear{van1998asymptotic}).

\hfill $\square$

\subsection{Proof of Proposition \ref{prop:etahatb_distr}}
\label{app:proof_prop2}
In order to use the mean value theorem on the relevant functions, and to ensure the results in \eqref{eqn:fi_xi_conv} and \eqref{eqn:fi_xi2_conv} hold also when using $g_\bt$, we further assume:
\begin{enumerate}[resume, label=\textbf{(A\arabic*)}, leftmargin=2em]
    \item The point $\bts$ is in the interior of the parameter space $\mB$.\label{assump:bts_interior}
    \item The signal density $f_s$ is continuously differentiable in $x$. \label{assump:fs_cont_diff}
\end{enumerate}

Set $\nu_\bt = (\tht_{0,\bt}, \del_{0,\bt})$ with
\[
\tht_{0,\bt} = \frac{\tht_\bt}{\|S_\bt\|_{G_\bt}} = \intmx S_{0,\bt}(x)dF(x), \quad
\del_{0,\bt} = \frac{\del_\bt}{\|S_\bt\|_{G_\bt}} = \intmx S_{0,\bt}(x)dF(x)
\]
where $S_{0,\bt}(x) = \Sd_{\bt}(x)/\|S_{\bt}\|_{G_\bt}$ and let its estimator be $\nuh_{\bt} = (\thth_{0, \bt}, \delh_{0, \bt})$ with
\[
\thth_{0,\bt} = \frac{\thth_{\bt}}{\|S_{\bt}\|_{G_\bt}} = \frac{1}{N}\sumk n_iS_{0,\bt}(x_i) \quad \text{and} \quad
\delh_{0,\bt} = \frac{\delh_{\bt}}{\|S_{\bt}\|_{G_\bt}} = \frac{1}{M}\sumk m_iS_{0,\bt}(x_i),
\]
such that, for any fixed $\bt \in \mB$, $\nuh_{\bt} \pconv \nu_{\bt}$.
Recall $\bts$ is the minimizer of the Kullback-Leibler divergence between $G_\bt$ and $F$. Hence, $\eta = W(\nu_{\bts})$ and $\widehat\eta_{\bth} = W(\nuh_{\bth})$, with $W$ defined in \eqref{eqn:T_def}. 

Applying the mean value theorem to \eqref{eqn:bt_score} gives us
\begin{equation}
\label{eqn:bth_bts_MVT}
    \begin{aligned}
        \frac{1}{T_b}\sumk m_i \Big[\pb \log g_{\bt,i} \Big]_{\bt = \bth}  =& 
        \frac{1}{T_b}\sumk m_i \Big[\pb\log g_{\bt, i}\Big]_{\bt = \bts} + \\ & 
        \Bigg[\frac{1}{T_b}\sumk m_i \Big[\pb^2  \log g_{\bt, i}\Big]_{\bt = \btt}\Bigg](\bth -\bts)\\
        \implies
        \sqrt{M}(\bth - \bts) =& \widehat{J}_{\btt}^{-1} \frac{1}{\sqrt{M}}\sumk m_i \Big[\pb \log g_{\bt, i}\Big]_{\bt = \bts},
    \end{aligned}
\end{equation}
where $\widehat{J}_\bt = \Big[-\frac{1}{M}\sumk m_i \pb^2  \log g_{\bt, i}\Big]$ and $\btt$ is a point lying between $\bth$ and $\bts$;  hence, $\btt \pconv \bts$. Similar arguments to those used in Section \ref{app:proof_bt_conv}, but applied to  $\partial^2_{\bt} \log g_{\bt,i}$, lead to  $\widehat{J}_{\bt}$ converging uniformly to $J_\bt = \mE_{F_b}\Big[-\pb^2 \log  g_{\bt}(Y)\Big]$ in probability, for any fixed $\bt \in \mB$ under \eqref{eqn:sparsness2}. Therefore,

 \begin{equation}
    \label{eqn:J_btt_pconv}
     \widehat{J}_{\widetilde{\beta}} \pconv J_{\bts}.
 \end{equation}

Next, using the mean value theorem on $\thth_{0,\bt}$ gives:
\begin{equation}
    \label{eqn:thth_exp_0}
    \begin{aligned}
        \thth_{0,\bth} =& \thth_{0,\bts} + \Big[\pb\thth_{0,\bt}\Big]_{\bt = \bt^\dag}^T(\bth - \bts)\\
        \sqrt{N}\Big(\thth_{0,\bth} - \tht_{0,\bts}\Big) = &
        \frac{1}{\sqrt{N}}\sumk n_i \Big(S_{0,\bts}(x_i) - \tht_{0,\bts}\Big) \\ & + \sqrt{\frac{N}{M}}\Big[\pb\thth_{0,\bt}\Big]_{\bt = \bt^\dag}^T\sqrt{M}(\bth - \bts)
    \end{aligned}
\end{equation}
where $\bt^\dag$ is a point lying between $\bth$ and $\bts$, leading to $\bt^\dag \pconv \bts$. Following arguments similar to those used to prove \eqref{eqn:J_btt_pconv} we get
\begin{equation}
    \label{eqn:pb_ththb_conv}
    \Big[\pb\thth_{0,\bt}\Big]_{\bt = \bt^\dag} \pconv \Big[\pb\tht_{0,\bt}\Big]_{\bt = \bts}.
\end{equation}
Combining \eqref{eqn:thth_exp_0} and \eqref{eqn:pb_ththb_conv} and setting $\xi_{1,i} = S_{0,\bts}(x_i) - \tht_{0,\bts}$, leads to:

\begin{equation}
    \label{eqn:thth_exp}
    \begin{aligned}
        \sqrt{N}\Big(\thth_{0,\bth} - \tht_{0,\bts}\Big) =& 
        \frac{1}{\sqrt{N}}\sumk n_i \xi_{1,i} + \sqrt{\frac{N}{M}}\Big[\pb\tht_{0,\bt}\Big]_{\bt = \bts}^T\sqrt{M}(\bth - \bts)
        + o_P(1).
    \end{aligned}
\end{equation}

A similar expansion can also be obtained for $\delh_{0,\bth}$ when setting $\xi_{2,i} = S_{0,\bts}(x_i) - \del_{0,\bts}$: 
\begin{equation}
    \label{eqn:delh_exp}
    \begin{aligned}
        \sqrt{M}\Big(\delh_{0,\bth} - \del_{0,\bts}\Big) =&
        \frac{1}{\sqrt{M}}\sumk m_i \xi_{2,i} + 
        \Big[\pb\del_{0,\bt}\Big]_{\bt = \bts}^T\sqrt{M}(\bth - \bts)+ o_P(1).
    \end{aligned}
\end{equation} 
Applying the mean value theorem on $W$ gives us:
\begin{equation}
    \label{eqn:etah_bt_MVT}
    \begin{aligned}
        &\sqrt{\frac{MN}{M+N}}\Big(\widehat\eta_{\bth} - \eta\Big) =
        \sqrt{\frac{MN}{M+N}}\Big(W\big(\nuh_{\bth}\big) - W\big(\nu_{\bts}\big)\Big) \\
        =&
        \sqrt{\frac{M}{M+N}}\pone W(\nut)\sqrt{N}\Big(\thth_{0,\bth} - \tht_{0,\bts}\Big) \\ &
        + 
        \sqrt{\frac{N}{M+N}}\ptwo W(\nut)\sqrt{M}\Big(\delh_{0,\bth} - \del_{0,\bts}\Big)
    \end{aligned}
\end{equation}
where $\nut$ is a point lying between $\nuh_{\bth}$ and $\nu_\bts$. From \eqref{eqn:thth_exp} and \eqref{eqn:delh_exp}, it follows that $\delh_{0,\bth}\pconv \del_{0,\bts}$, $\thth_{0,\bth}\pconv \tht_{0,\bts}$, $\nuh_{\bth} \pconv \nu_{\bts}$ and therefore $\nut \pconv \nu_{\bts}$. Combining \eqref{eqn:bth_bts_MVT}, \eqref{eqn:thth_exp} , \eqref{eqn:delh_exp} and \eqref{eqn:etah_bt_MVT} we obtain
\begin{equation}
\label{eqn:etah_expansion_xi}
    \begin{aligned}
        \sqrt{\frac{MN}{M+N}}\Big(\widehat\eta_{\bth} - \eta\Big) =& 
    \sqrt{\frac{M}{M+N}}\pone W(\nu_{\bts})\frac{1}{\sqrt{N}}\sumk n_i \xi_{1,i} \\
    & 
    + \sqrt{\frac{N}{M+N}}\frac{1}{\sqrt{M}}\sumk m_i \xi_{3,i} + o_P(1)
    \end{aligned}
\end{equation}
in which
\begin{equation}
    \begin{aligned}
        \xi_{3,i} =& \ptwo W(\nu_{\bts})\xi_{2,i} + 
        \Gamma_{\bts}^T J_{\bts}^{-1}\big[\pb\log g_{\bt,i}\big]_{\bt = \bts} \quad \text{and}\\
        \Gamma_{\bts} =& \Bigg(
        \pone W(\nu_{\bts})\Big[\pb\tht_{0,\bt}\Big]_{\bt = \bts} + 
        \ptwo W(\nu_{\bts})\Big[\pb\del_{0,\bt}\Big]_{\bt = \bts}
        \Bigg).\\
    \end{aligned}
\end{equation}
Since the search region $\X$ is compact, assumptions \ref{assump:gb_cont_bt} - \ref{assump:mB_compact} ensure the boundedness of the relevant functions and consequently the boundedness of the $\xi_{1,i}$'s and $\xi_{2,i}$'s. From analogous arguments to those used in proving \eqref{eqn:fi_xi_conv}, \eqref{eqn:fi_xi2_conv}, and \eqref{eqn:J_btt_pconv}, we obtain:
\begin{equation}
    \label{eqn:xiji_conv}
    \begin{aligned}
        &\sqrt{T}\sumk f_i\xi_{1,i} \to 0, \quad  \sqrt{T_b}\sumk b_i\xi_{2,i} \to 0
        \\
        &\sumk f_i\xi_{1,i}^2 \to \sig_{\bts, \tht_0}^2 = \frac{\sig_{\bts, \tht}^2}{\|S_{\bts}\|_{G_{\bts}}^2}, 
        \quad
        \sumk b_i\xi_{2,i}^2 \to \sig_{\bts, \del_0}^2 = \frac{\sig_{\bts, \del}^2}{\|S_{\bts}\|_{G_{\bts}}^2}
    \end{aligned}
\end{equation}
with 
\[
\sig_{\bts, \tht}^2 = \intmx \big(\Sd_{\bts}(x)\big)^2dF(x) - \tht_{\bts}^2, \quad 
\sig_{\bts, \del}^2 = \intmx \big(\Sd_{\bts}(x)\big)^2dF_b(x) - \del_{\bts}^2
\]
which leads to
\begin{equation}
    \label{eqn:bi_x2i_conv}
        \sqrt{T_b}\sumk b_i\xi_{3,i} \to 0 \quad \text{and}\quad \sumk b_i \xi_{3,i}^2  \to \Lambda_{\bts}
\end{equation}
where 
\begin{equation}
    \label{eqn:lambda_bts_def}
    \begin{aligned}
        \Lambda_{\bts} 
        =& \ptwo W(\nu_{\bts})^2 \sig_{\bts,\delta_0}^2 +  \Gamma_{\bts}^T J_{\bts}^{-1}V_{\bts} J_{\bts}^{-1}\Gamma_{\bts} + 
        2 \frac{\ptwo W(\nu_{\bts})\Gamma_{\bts}^T J_{\bts}^{-1}}{\|S_{\bts}\|_{G_{\bts}}}C_{\bts}\\
        =&
        \frac{\sig_{\bts,\delta}^2(\tht_{\bts}-\|S_{\bts}\|_{G_{\bts}})^2}{(\|S_{\bts}\|_{G_{\bts}}- \del_{\bts})^4}  +  \Gamma_{\bts}^T J_{\bts}^{-1}V_{\bts} J_{\bts}^{-1}\Gamma_{\bts}\\ &
        + 
        2 \frac{(\tht_{\bts}-\|S_{\bts}\|_{G_{\bts}})\Gamma_{\bts}^T J_{\bts}^{-1}}{(\|S_{\bts}\|_{G_{\bts}}- \del_{\bts})^2}C_{\bts}
    \end{aligned}
\end{equation}
with
\begin{align*}
    V_{\bts} = Var_{F_b}\Bigg(\Big[\pb \log g_{\bt}(Y)\Big]_{\bt = \bts}\Bigg) 
    =
    \intmx \Big[\pb \log g_{\bt}(y)\Big]_{\bt = \bts}\Big[\pb \log g_{\bt}(y)\Big]_{\bt = \bts}^TdF_b(y)
\end{align*}
and
\begin{align*}
    C_{\bts} = Cov_{F_b}\Bigg(\Big[\pb \log g_{\bt}(Y)\Big]_{\bt = \bts}, \Sd_{\bts}(Y)\Bigg)
    = \intmx \Sd_{\bts}(y)\Big[\pb \log g_{\bt}(y)\Big]_{\bt = \bts} dF_b(y).
\end{align*}

Combining \eqref{eqn:etah_expansion_xi}, \eqref{eqn:xiji_conv}, \eqref{eqn:bi_x2i_conv} and using Lemma \ref{prop:lemma1} leads to
\begin{equation}
    \label{eqn:etahb_dist_sig_bts}
    \sqrt{\frac{MN}{M+N}}(\widehat\eta_{\bth} - \eta) \dconv \N(0, \sig_{\bts,\eta}^2)
\end{equation}
with 
\begin{equation}
    \label{eqn:sig_eta_bts_def}
    \sig_{\bts,\eta}^2 =  \frac{(1-\pi)\sig_{\bts, \tht}^2}{(\|S_{\bts}\|_{G_{\bts}} - \del_{\bts})^2}  + \pi \Lambda_{\bts}.
\end{equation}

Consider the following consistent estimators for the terms involved in $\sig_{\bts,\eta}^2$:
\begin{align*}
    & 
    \|S_{\bth}\|_{G_{\bth}} \pconv \|S_{\bts}\|_{G_{\bts}};\\ &
    \sigh_{\bth, \tht}^2 = \frac{1}{N}\sumk n_i \Sd_{\bth}(x_i) - \thth_{\bth}^2 \pconv \sig_{\bts, \tht}^2;\\ &
    \sigh_{\bth, \del}^2 = \frac{1}{M}\sumk m_i \Sd_{\bth}(x_i) - \delh_{\bth}^2 \pconv \sig_{\bts, \del}^2;\\ &
    \big[\pb \thth_{0,\bt}\big]_{\bt = \bth} \pconv 
    \big[\pb \tht_{0,\bt}\big]_{\bt = \bts};\quad 
    \big[\pb \delh_{0,\bt}\big]_{\bt = \bth} \pconv 
    \big[\pb \del_{0,\bt}\big]_{\bt = \bts};\\ &
    \widehat\Gamma_{\bth} = \Big[\pone W(\nuh_{\bth})\big[\pb \thth_{0,\bt}\big]_{\bt = \bth} +
    \ptwo W(\nuh_{\bth})\big[\pb \delh_{0,\bt}\big]_{\bt = \bth}
    \Big] \pconv \Gamma_{\bts}; \\ &
    \widehat{J}_{\bth} \pconv J_{\bts}; \\ &
    \widehat{C}_{\bth} = \frac{1}{M}\sumk m_i \Sd_{\bth}(x_i)\big[\pb \log g_{\bt}(x_i)\big]_{\bt = \bth} \pconv C_{\bts} ;\\ &
    \widehat{V}_{\bth} = \frac{1}{M}\sumk m_i \big[\pb \log g_{\bt}(x_i)\big]_{\bt = \bth}\big[\pb \log g_{\bt}(x_i)\big]_{\bt = \bth}^T \pconv V_{\bts}; 
\end{align*}
which can be shown to be consistent using the same argument used to show validity of \eqref{eqn:J_btt_pconv}. A consistent estimator, $\sigh_{\bth,\etah}^2$ , for $\sig_{\bts,\eta}^2$  can be obtained by plugging in the estimators above in place of their limit in \eqref{eqn:lambda_bts_def} and \eqref{eqn:sig_eta_bts_def}. 

Finally, dividing the left hand side of \eqref{eqn:etahb_dist_sig_bts} by $\sigh_{\bth,\etah}$ and applying Slutsky's theorem gives us the desired result.

\hfill $\square$

\subsection{Proof of Proposition \ref{prop:thth_0bts_distr}}
\label{app:proof_prop3}
Recall that $\bt = (\al, \lambda) \in \mB \subset \R^p$. Let $\partial$ and $\partial^2$ denote, respectively, the $p-1$ dimensional gradient vector and the $(p-1) \times (p-1)$ Hessian matrix of $q_\al$, with derivatives taken respect to $\al$, i.e., the first $p-1$ dimensional component of $\bt$. Similar to \eqref{eqn:MLE}, by applying the mean value theorem on the score function of $q_\al$, we obtain:
\begin{equation}
    \label{eqn:alh_expansion_0}
        \begin{aligned}
            \sumk n_i \Big[\partial \log q_{\al, i}\Big]_{\al = \alh} =& \sumk n_i \Big[\partial \log q_{\al, i}\Big]_{\al = \als} + \Bigg[\sumk n_i \Big[\partial^2 \log q_{\al,i}\Big]_{\al = \tilde{\al}}\Bigg] (\alh - \als)\\
            \implies
            \sqrt{N}(\alh-\als) =&
            \widehat{\mJ}_{\tilde\al}^{-1}
            \frac{1}{\sqrt{N}}\sumk n_i 
            \Big[\partial \log q_{\al,i}\Big]_{\al = \als}
        \end{aligned}
\end{equation}
where, $\widehat\mJ_{\al} = \Big[-\frac{1}{N}\sumk n_i \partial^2 \log q_{\al,i}\Big]$, $q_{\al, i} = \int_{\Delta_i} q_{\al}(x)dx$, and $\widetilde\al$ is a point lying between $\als$ and $\alh$. Similar arguments presented in Section \ref{app:proof_bt_conv} lead to $\alh \pconv \als$ and, therefore, $\widetilde\al \pconv \als$. Analogous to \eqref{eqn:J_btt_pconv}, we have 
\begin{equation}
    \label{eqn:mJ_hat_alt_conv}
    \widehat\mJ_{\widetilde\al} \pconv \mJ_{\als} = \mE_{F}\Big(-\Big[\partial^2\log q_{\al}(X)\Big]_{\al = \als}\Big).
\end{equation}
Combining \eqref{eqn:mJ_hat_alt_conv} and \eqref{eqn:alh_expansion_0} yields:
\begin{equation}
    \label{eqn:alh_expansion}
    \sqrt{N}(\alh-\als) =
            \mJ_{\als}^{-1}
            \frac{1}{\sqrt{N}}\sumk n_i \Big[\partial \log q_{\al,i}\Big]_{\al = \als} + o_P(1).
\end{equation}

Following similar arguments used to derive \eqref{eqn:pb_ththb_conv}, the mean value theorem, and \eqref{eqn:alh_expansion} yield:
\[
    \thth_{0,\bth} = \thth_{0,\bts} + D_{\als}^T(\alh - \als) + o_P(1)
\]
with  $D_{\als} = \Big[\partial \tht_{0,\bt}\Big]_{\al = {\als}}$. Hence,
\begin{equation}
    \label{eqn:thth_alh_expansion}
    \begin{aligned}
        \sqrt{N}(\thth_{0,\bth} - \tht_{0,\bts}) =&
        \frac{1}{\sqrt{N}}\sumk n_i \Big(S_{0,\bts}(x_i) - \tht_{0,\bts}\Big) + D_{\als}^T\sqrt{N}(\alh - \als) + o_P(1) \\ 
        =&
        \frac{1}{\sqrt{N}}\sumk n_i \xi_i + o_P(1)
    \end{aligned}
\end{equation}
where 
$\xi_i = \Big(S_{0,\bts}(x_i) - \tht_{0,\bts}\Big) + D_{\als}^T\mJ_{\als}^{-1}\partial \log q_{\al,i}$. Once again, the argument used in  \eqref{eqn:xiji_conv} - \eqref{eqn:bi_x2i_conv} leads to $\xi_i$'s being bounded, $\sqrt{T}\sumk f_i\xi_i \to 0$ , and
\begin{equation}
    \label{eqn:fixi2_conv}
    \sumk f_i \xi_i^2 \to \sig_{\bts,\tht_0}^2 = \frac{\sig_{\bts, \tht}^2}{\|S_{\bts}\|_{G_{\bts}}^2} + D_{\als}^T\mJ_{\als}^{-1}\mV_{\als}\mJ_{\als}^{-1}D_{\als} + 2 D_{\als}^T\mJ_{\als}^{-1}\mC_{\als},
\end{equation}
in which 
\[
\mV_\als = Var_{F}\Big(\big[\partial \log q_\al (X)\big]_{\al = \als}\Big) = \intmx \big[\partial \log q_\al (X)\big]_{\al = \als}\big[\partial \log q_\al (X)\big]_{\al = \als}^TdF(x)
\]
and
\[
\mC_{\als} = Cov_{F}\Big(S_{0,\bts}(X), \big[\partial \log q_\al (X)\big]_{\al = \als}\Big) = \intmx S_{0,\bts}(x)\big[\partial\log q_\al (x)\big]_{\al = \als}dF(x)
\]
when \eqref{eqn:sparsness} holds. Therefore, from Lemma \ref{prop:lemma1}, we have
\begin{equation}
    \label{eqn:thth_bth_dist_0}
    \sqrt{N}\Big(\thth_{0,\bth} - \tht_{0,\bts}\Big) \dconv 
        \N\Big(0, \sig_{\bts,\tht_0}^2\Big).
\end{equation}

A consistent estimator for $\sig_{\bts,\tht_0}^2$, $\sigh_{\bth,\thth_0}^2$,  can be obtained by replacing the elements in \eqref{eqn:fixi2_conv} with the corresponding consistent estimators:

\begin{align*}
    &
    \|S_{\bth}\|_{G_{\bth}} \pconv \|S_{\bts}\|_{G_{\bts}};\\ &
    \sigh_{\bth, \thth}^2 = \frac{1}{N}\sumk n_i \Sd_{\bth}(x_i) - \thth_{\bth}^2 \pconv \sig_{\bts, \tht}^2;\\ &
    \widehat{D}_{\alh} = \big[\partial \thth_{0,\bt}\big]_{\al = \alh} \pconv 
    \big[\partial \tht_{0,\bt}\big]_{\al = \als} = D_{\als};\\ &
    \widehat{\mJ}_{\alh} \pconv \mJ_{\als}; \\ &
    \widehat{\mC}_{\alh} = \frac{1}{N}\sumk n_i S_{0,\bth}(x_i)\big[\partial \log q_{\al}(x_i)\big]_{\al = \alh} \pconv \mC_{\als}; \\ &
    \widehat{\mV}_{\alh} = \frac{1}{N}\sumk n_i \big[\partial \log q_{\al}(x_i)\big]_{\al = \alh}\big[\partial \log q_{\al}(x_i)\big]_{\al = \alh}^T \pconv \mV_{\als}. \\
\end{align*}

Finally, dividing the left hand side of \eqref{eqn:thth_bth_dist_0} by $\sigh_{\bth,\thth_0}$ and applying Slutsky's theorem gives the desired result.

\end{document}